\documentclass[twocolumn,showpacs,preprintnumbers,amsmath,amssymb,prb]{revtex4-1}
\usepackage{graphicx}
\usepackage{amsmath, amsthm, amssymb}
\usepackage{latexsym}
\usepackage{amssymb}
\usepackage{bm}
\usepackage{dcolumn}
\usepackage{epstopdf}
\usepackage{color}
\usepackage{braket}
\usepackage{filecontents}

\newcommand{\UP}{UPt$_3$}
\newcommand{\SHE}{superfluid $^3$He}

\begin{document}

\title{Spin Susceptibility of the Topological Superconductor \UP\ from Polarized Neutron Diffraction }
\date{\today}

\author{W. J. Gannon}
\altaffiliation{Present address: Department of Physics, Texas A\&M University, TX 77843 USA}
\affiliation{Department of Physics and Astronomy, Northwestern University, Evanston, IL 60660 USA}

\author{W. P. Halperin}
\affiliation{Department of Physics and Astronomy, Northwestern University, Evanston, IL 60660 USA}

\author{M. R. Eskildsen}
\affiliation{Department of Physics, University of Notre Dame, Notre Dame, IN 46556 USA}

\author{Pengcheng Dai}
\affiliation{Department of Physics and Astronomy, Rice University, Houston, TX 77005 USA}

\author{U. B. Hansen}
\affiliation{Niels Bohr Institute, University of Copenhagen, DK-2100 Copenhagen, Denmark}

\author{K. Lefmann}
\affiliation{Niels Bohr Institute, University of Copenhagen, DK-2100 Copenhagen, Denmark}

\author{A. Stunault}
\affiliation{Institut Laue Langevin, 71 Avenue des Martyrs, F-38042 Grenoble, France}

\begin{abstract}
Experiment and theory indicate that \UP\ is a topological superconductor in an odd-parity state, based in part from temperature independence of the NMR Knight shift.  However,  quasiparticle spin-flip scattering near a surface, where the Knight shift is measured, might be responsible. We use polarized neutron scattering to measure the bulk susceptibility with $H||c$, finding consistency with the Knight shift but inconsistent with theory for this field orientation. We infer that neither spin susceptibility nor Knight shift are a reliable indication of odd-parity.
\end{abstract}

\maketitle

Broad interest in topological superconductivity stems in part from the recognition that this class of superconductors can host Majorana fermionic excitations  of special importance for quantum information processing.~\cite{Bee2013} The first discovery of such a material in 1972 was \SHE, an odd-parity, Cooper paired superfluid with a  $k_x \pm i k_y$ $p$-wave orbital order parameter.~\cite{Osh1972a,Osh1972b}
More than a decade later, evidences for two other three-dimensional quantum  materials, \UP~\cite{Stewart_PRL_1984,Joynt_RevModPhys_2002} and Sr${_2}$RuO${_4}$,~\cite{Mae1994,Mac2003} were reported. These are now thought to be topological superconductors which in this context means they have odd-parity, chiral symmetry and broken time reversal symmetry that should host Majorana zero modes in vortex cores.~\cite{Iva2001}
Support  for  odd-parity comes in part from the temperature independence of the NMR Knight shift, taken to be proportional to the spin susceptibility.  This  was  observed for all magnetic field orientations,~\cite{Ish1998, Tou1996a, Tou1998b}   as is the case for the A-phase of \SHE.  However, temperature independence of the spin susceptibility for magnetic fields along the $c$-axis of \UP~\cite{Tsu2012} is only possible if the spin-orbit interaction is negligible, contrary to expectations.~\cite{Sau1994,Graf_PRB_2000}
Since the Knight shift is measured only within a London penetration depth of the surface it may not reflect bulk behavior motivating our investigation of the bulk magnetic susceptibility using spin-polarized neutron diffraction.

A remarkable parallel between \UP\ and \SHE\ is that both exhibit multiple superconducting phases.  Based in part on the field-temperature phase diagram for \UP,\cite{Ade1990} Sauls proposed a $f$-wave order parameter of the form  $k_z(k_x \pm i k_y)^2$ with E$_{2u}$
 symmetry,~\cite{Sau1994} A degeneracy in the manifold of possible $f$-wave states is lifted by a symmetry breaking field (SBF), thereby giving rise to three thermodynamically stable superconducting phases A, B, and C, Fig.~1.
The origin of the SBF was associated~\cite{Sau1994} with antiferromagnetism that onsets at a temperature of ~6 K,~\cite{Aeppli_PRL_1988} an order of magnitude higher than the superconducting transition of the pure crystal, $T_c = 0.56$\,\,K.~\cite{Kycia_PRB_1998} Recent neutron scattering experiments show that  the antiferromagnetism is not static and that most likely the antiferromagnetic fluctuations  are responsible for the SBF.~\cite{Gan2013,Carr2017}  For \SHE, 
strong coupling of quasiparticle interactions at pressures above the polycritical point at $P=21$\,bar  correspond to the SBF that stabilizes the chiral A-phase with respect to the fully gapped B-phase.~\cite{Leg1975,Whe1975}

%**************************************************************************************************************************
%************************************Figure 1************************************************
%**************************************************************************************************************************
\begin{figure}
\includegraphics[width=80mm]{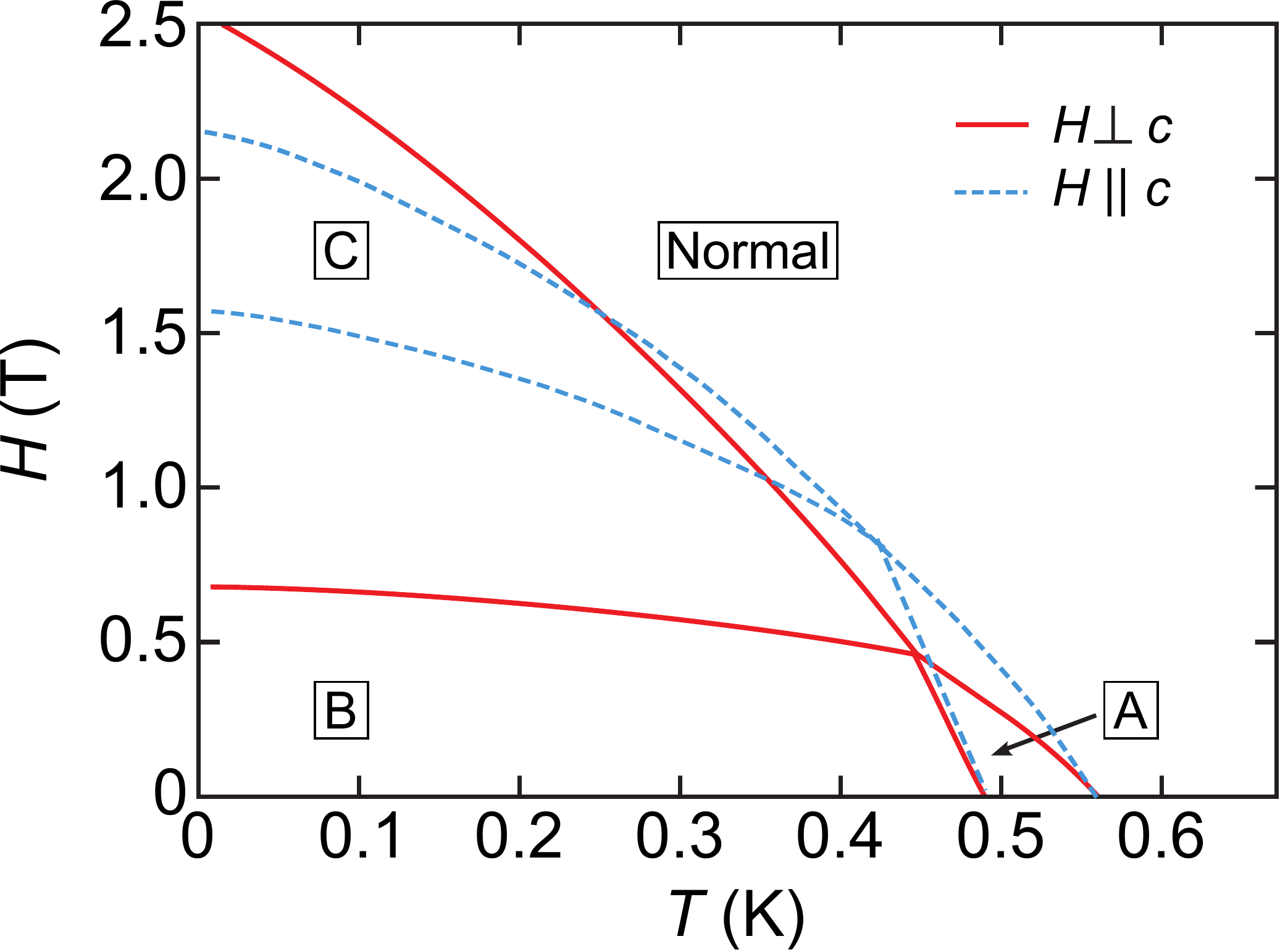}
\caption{\label{PhaseDiagram} (Color online)  Schematic of the phase diagram of \UP\ showing the superconducting A, B, and C phases for magnetic fields perpendicular (and parallel) to the $c$ axis, red (blue dashed) curves, taken in part from Adenwalla {\it et al.}~\cite{Ade1990} adjusted at $H = 0$ and for $H(T \approx 50$~mK) from measurements on the high purity crystals in this and earlier work.~\cite{Kycia_PRB_1998}}
\end{figure}

The most recent experimental evidence for the assignment of E$_{2u}$ symmetry to the order parameter for \UP\ comes from phase-sensitive measurements of Josephson tunneling in the B-phase~\cite{Strand_PRL_2009} and the directional tunneling study of nodal structure in the A-phase.~\cite{Strand_Science_2010}  Moreover, polar Kerr effect experiments~\cite{Schemm2014} have identified the B-phase as breaking time-reversal symmetry, absent in the A-phase, consistent with the E$_{2u}$ theory of Sauls~\cite{Sau1994} for which only the B-phase has chiral symmetry.  These results are complemented by measurements of the linear temperature dependence of the London penetration depth from magnetization~\cite{Sig1995,Sch1999} and small angle neutron scattering~\cite{Gan2015} consistent with the nodal structure of the energy gap in this theory.  On the other hand, measurements of the angle dependence of  thermal conductivity~\cite{Mach2012} have been interpreted in terms of a time reversal symmetric, $f$-wave order parameter structure with E$_{1u}$ symmetry, at odds with the other experimental work listed above. Nonetheless, the preponderance of evidence favors the E$_{2u}$ symmetry assignment.  

The motivation for the present work is to investigate the spin susceptibility, and the importance of the spin-orbit interaction, since this bears on the question of the assignment of odd-parity to \UP.  An odd-parity superconductor should be in a  spin-triplet state. In the presence of a significant spin-orbit interaction, it is natural to take the direction for orbital quantization parallel to the $c$-axis for both \UP\ and Sr${_2}$RuO${_4}$, forcing  the spin quantization axis to be perpendicular, {\it i.e.} in the $ab$-plane.  As a consequence, for magnetic fields $H$ in the basal plane, the spin susceptibility will be temperature independent, and equal to the normal state susceptibility.  However, with $H||c$ it would be expected to be strongly temperature dependent,  similar to the behavior of the susceptibility in a  spin-singlet state of a conventional superconductor, approaching zero at low temperatures.  This is not observed in the  NMR Knight shift experiments.  Rather,  the Knight shift for \UP\ was found to be temperature independent, irrespective of orientation of the magnetic field relative to the crystal.~\cite{Tou1996a,Tou1998b} We note that in the case of Sr$_2$RuO$_4$ measurements of the spin susceptibility in the superconducting state have only been made with the magnetic field in the $ab$-plane~\cite{Ish1998,Ish2001, Duffy_PRL_2000} owing to an extremely small upper critical field for $H||c$, $\approx 0.1$\,T.~\cite{Ish2001} 

The temperature independence of the Knight shift of \UP\ for all orientations of the magnetic field can only be accounted for theoretically by assuming a weak spin-orbit interaction.  This was proposed by Ohmi and Machida {\it et al.},~\cite{Ohmi_JPSJ_1996,Mach1999}  although it  seems unlikely given the large atomic mass  of the constituents in \UP\ and concomitant strength of spin-orbit coupling.  Moreover, the observed Pauli limiting of the upper critical field,~\cite{Shivaram_PRL_1986,Joynt_RevModPhys_2002} selectively for $H||c$, indicates that the orbital quantization axis is locked to the $c$-axis by a significant spin-orbit interaction.~\cite{Cho1991, Sau1994, Graf_PRB_2000}   In this case,  the interpretation of the NMR Knight shift must be incorrect.   One possibility is that  for superconductors with spin-orbit interaction, spin-flip scattering from the surface might account for the temperature independence of the $^{195}$Pt Knight shift,~\cite{Abr1962}  which is measured only within a London penetration depth of the surface ($\lambda_c \sim 4,000$~\AA, $\lambda_{ab} \sim 7,000$~\AA).~\cite{Gan2015}    Consequently, it is important to determine the spin susceptibility by other means, such as  polarized neutrons that penetrate into  the bulk superconductor with only modest attenuation of the beam intensity.  This is the approach we take here. 

In previous work we  measured the magnetic susceptibility with $H \perp c$ using polarized neutron scattering for a \UP\ crystal with high purity, residual resistance ratio (RRR) of 600 and mass 15~g.~\cite{Gan2012}
We found that the spin susceptibility is temperature independent in the superconducting state for this orientation consistent with Knight shift results.\cite{Tou1996a, Tou1998b} Here we extend these measurements for $H || c$ with a smaller crystal, 0.4~g, of even higher purity grown at Northwestern University (RRR = 900),~\cite{Gan2013} for which a strong temperature dependence of the susceptibility is expected from the E$_{2u}$ theory for odd-parity superconductivity with angular momentum locked to the $c$-axis.~\cite{Sau1994}
%**************************************************************************************************************************
%************************************Figure 2************************************************
%**************************************************************************************************************************
\begin{figure}
\includegraphics[width=79mm]{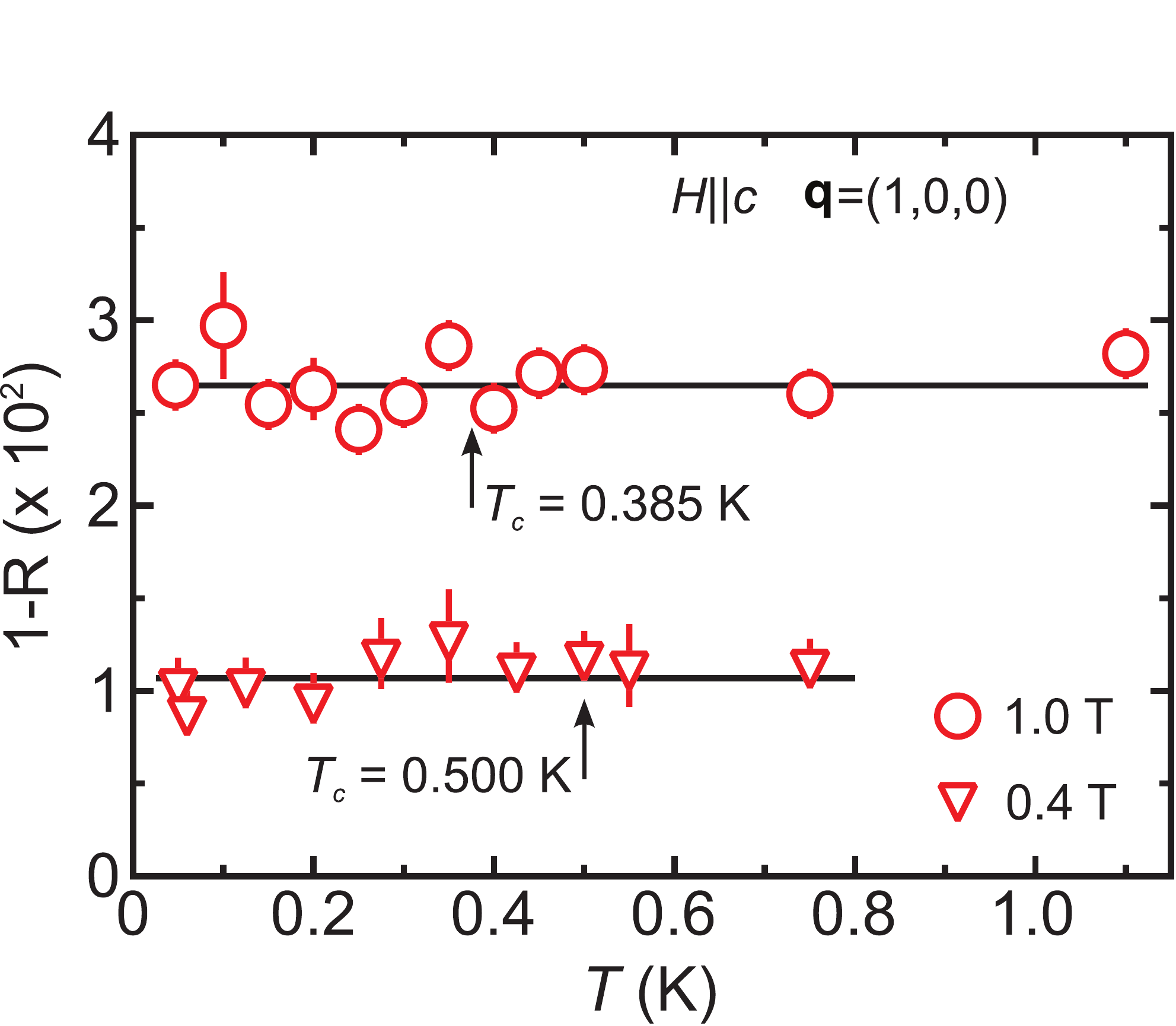}
\caption{\label{R}(Color online) One minus the flipping ratio $R$ as a function of temperature for the nuclear Bragg reflection (1,0,0) at several magnetic fields:  0.4 T (triangles), and 1.0 T (circles).  Black lines show the average value of 1--$R$ at each field.}  
\end{figure}

Neutron scattering experiments were conducted on the polarized neutron 2-axis diffractometer at the Institut Laue Langevin.~\cite{dir-107}
The bulk magnetization was determined by measuring the neutron spin-flipping ratio $R$ at the (1,0,0) nuclear Bragg reflection as described in more detail in previous work.~\cite{Gan2012,Gan2013} Here, $R$ is defined as the ratio of scattering cross sections 
for incident neutrons with spins parallel and antiparallel to the  applied magnetic field, and an arbitrary final spin state in each case.~\cite{Squires_book_1978} This procedure was pioneered by Shull and Wedgwood
for V$_3$Si,\cite{Shull_PRL_1966} and has been used to investigate the spin susceptibility of \UP\ and Sr$_2$RuO$_4$ for $H\perp c$.\cite{Stassis_PRB_1986,Duffy_PRL_2000,Gan2015} 

The flipping ratio can be expressed as,~\cite{Squires_book_1978}
\begin{equation} \label{R2}
1-R = \frac{2\gamma r_{0}}{\mu_{B}}\frac{M_{||}({\bf q})}{F_N({\bf q})},
\end{equation}

\noindent
in the limit $R\lesssim $\,1.  The nuclear structure factor is $F_N({\bf q})$ for a reflection being measured at wave vector ${\bf q}$ and  $M_{||}({\bf q})$ is the corresponding Fourier component of magnetization parallel to the applied field. Here $\gamma$ is the neutron gyromagnetic ratio, $r_0$ is the classical radius of the electron, and $\mu_{B}$ the Bohr magneton.
The (1,0,0)  reflection is ideally suited for measurements of the flipping ratio, because it is a relatively weak nuclear reflection and has the smallest possible ${\bf q}$ for the present scattering configuration, maximizing $1-R$ in Eq.~\ref{R2}. 

%**************************************************************************************************************************
%************************************Figure 3************************************************
%**************************************************************************************************************************

\begin{figure}
\includegraphics[width=80mm]{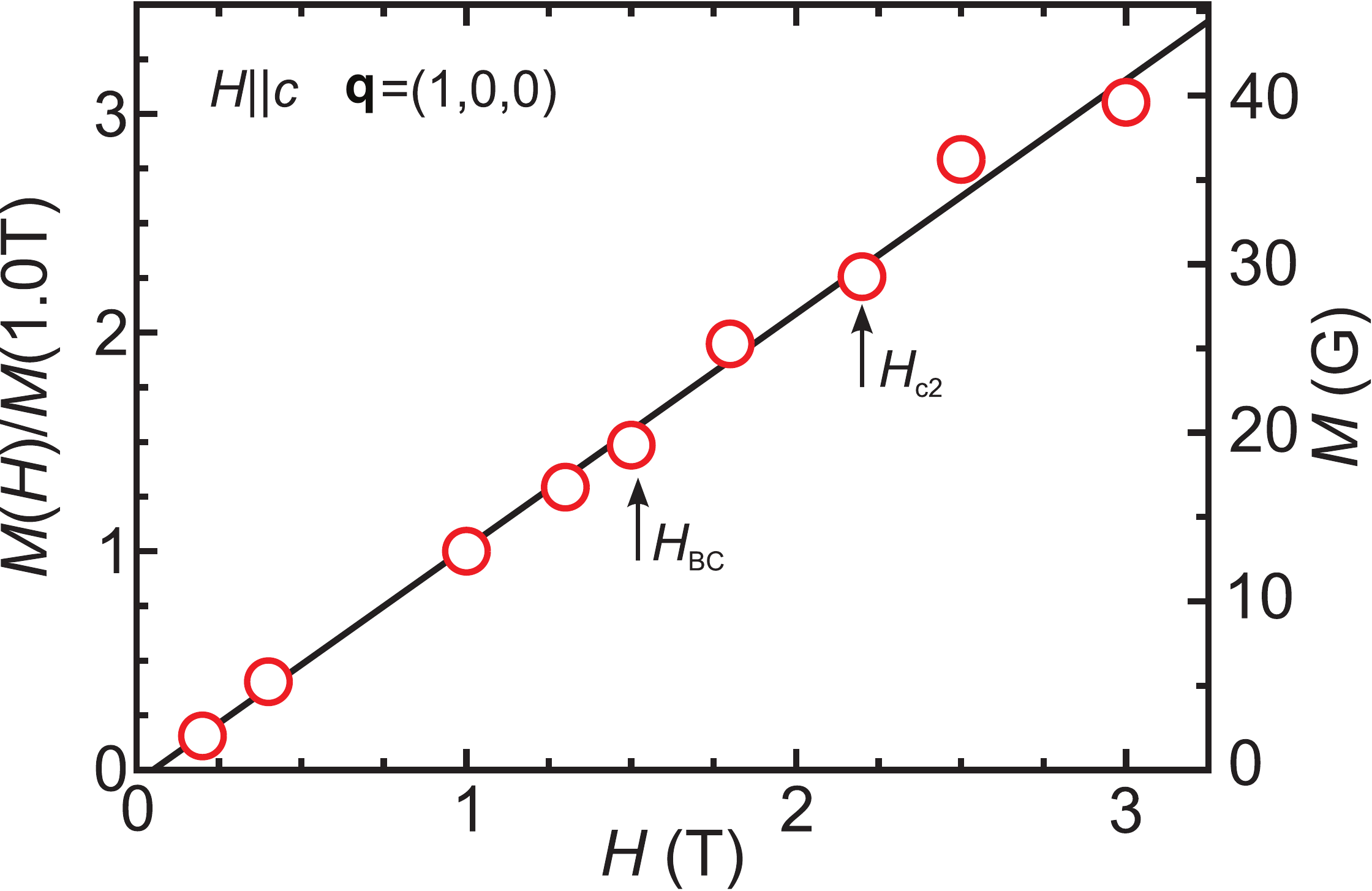}
\caption{\label{Mbar}(Color online)  Magnetization as a function of  magnetic field. At 0.4~T and 1.0~T where the temperature dependence was measured we show the average  for $T < T_c$; for all other fields the results are for $T \approx 50$~mK.
The proportionality to magnetic field indicates that the susceptibility can be taken to be $\chi = M/B$, and that the diamagnetic magnetization from the Meissner screening in the superconducting state is negligible.}
\end{figure}

The results at two magnetic fields, $H = 0.4$\,T and 1.0\,T, are shown in Fig.~\ref{R}. Within the statistical error of the measurement, the flipping ratio is found to be temperature independent. This indicates that the magnetization along the crystal $c$-axis does not change with temperature across the superconducting-normal transition, or for transitions between the different superconducting phases.

In order to determine the susceptibility from the magnetization, $M$, we examined its field dependence shown in Fig.~\ref{Mbar} normalized to the value at 1.0~T. This is found to be strictly proportional to field. Consequently the magnetic susceptibility can be taken to be $\chi = M/B$. The proportionality of the magnetization with field indicates that the non-linear diamagnetism from the Meissner screening currents in the superconducting state is insignificant.   This is different from other compounds like V$_3$Si or Sr$_2$RuO$_4$ for which polarized neutrons have been used to explore the superconducting state.~\cite{Shull_PRL_1966, Duffy_PRL_2000}  These materials have a significant contribution from diamagnetism which must be considered.  We have calculated the  diamagnetism  from Ginzburg-Landau theory~\cite{Brandt_PRB_2003} and find that in \UP\ it is negligible, consistent with the linear behavior in Fig.~\ref{Mbar}.
%**************************************************************************************************************************
%************************************Figure 4************************************************
%**************************************************************************************************************************

\begin{figure}
\includegraphics[width=80mm]{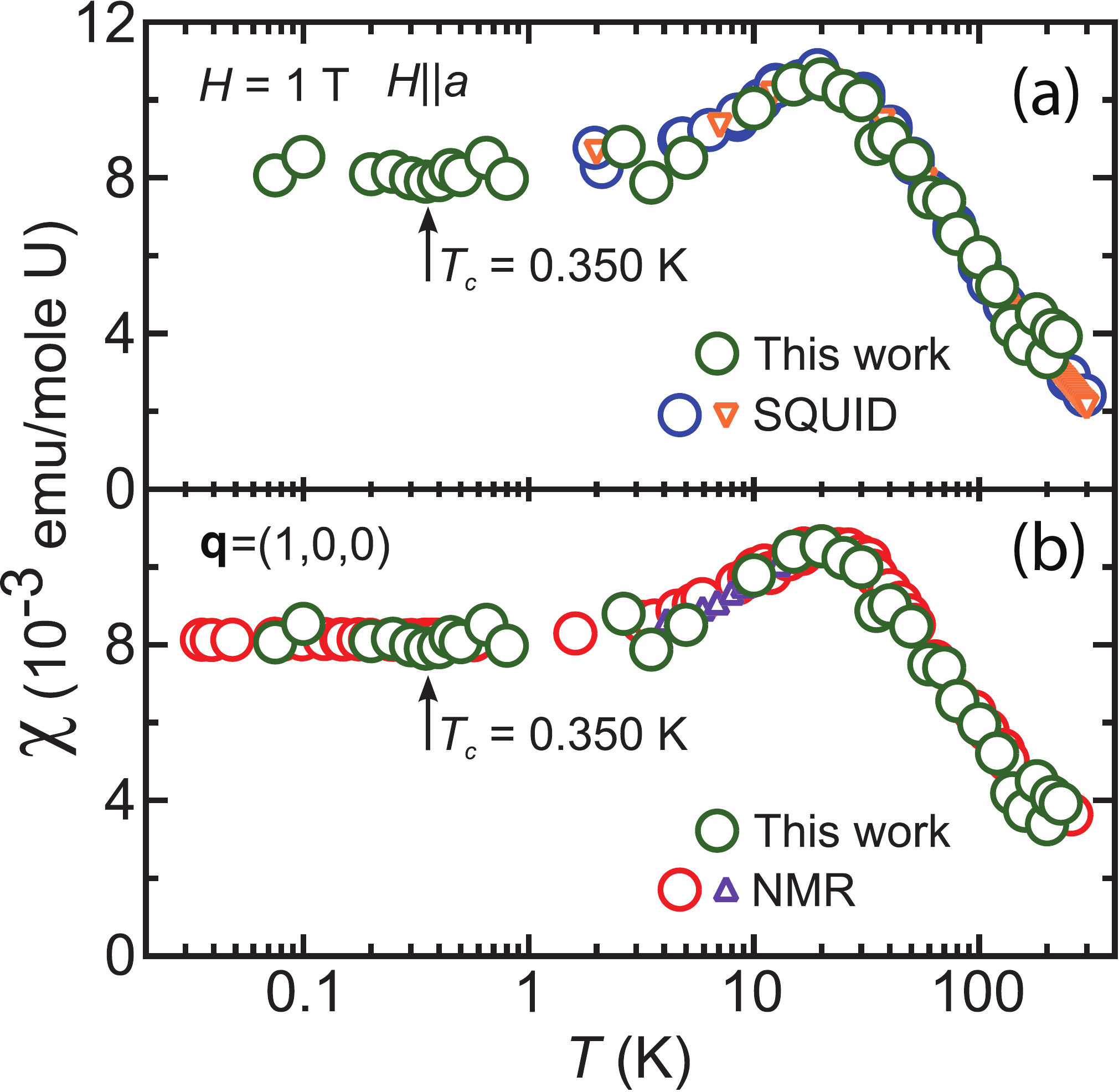}
\caption{\label{Chi_ab}(Color online) [Figure taken from Reference~\citenum{Gan2012}.] (a) Magnetic susceptibility measured by polarized neutron diffraction (solid green circles) for $H\perp c$,~\cite{Gan2012} SQUID measurements (open orange triangles)~\cite{Gan2012} and  susceptibility measurements (open blue circles).\cite{Frings_JMMM_1983}
(b) Susceptibility calculated from the Knight shift (open red circles~\cite{Tou1996a, Koh1990} and open purple triangles~\cite{Lee1993}).
The statistical error for the susceptibility measured by neutron scattering is approximately the size of the green circles.}
\end{figure}

%**************************************************************************************************************************
%************************************Figure 5************************************************
%**************************************************************************************************************************

\begin{figure}
\includegraphics[width=80mm]{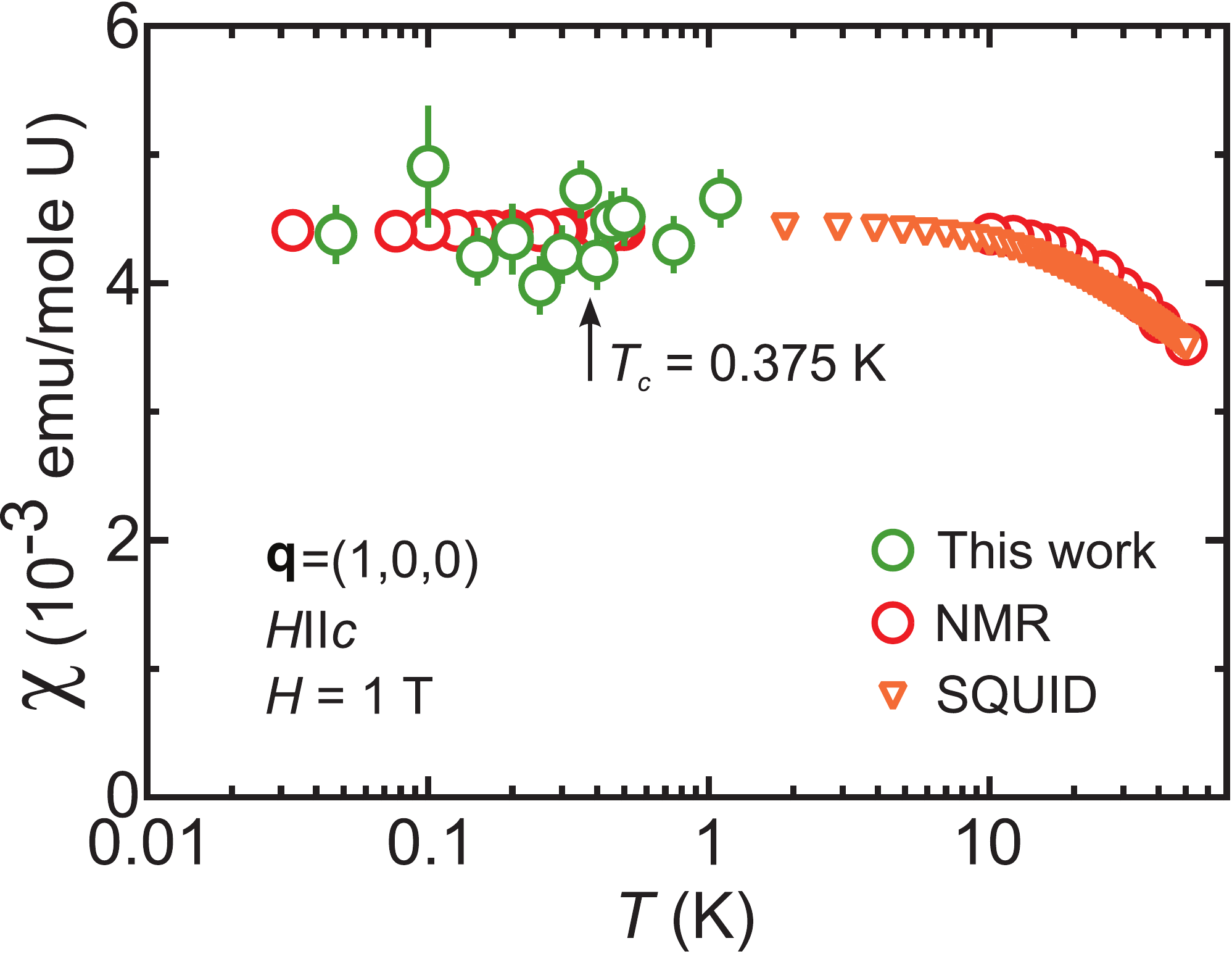}
\caption{\label{Chi_c}(Color online) Magnetic susceptibility measured by polarized neutron diffraction  for $H||c$ (solid green circles). The symbol notation is the same as for Fig.~\ref{Chi_ab}.}
\end{figure}

In principle we can determine the absolute magnetization directly from our measurements using Eq.~\ref{R2}.  However, this neglects potential complications such as extinction, neutron absorption, and depolarization of the incident neutron beam, in addition to the inherent ${\bf q}$ dependence which reflects the electronic orbital configuration contributing to the underlying magnetism being probed.~\cite{Squires_book_1978}  In a system with local moment magnetic character, the ${\bf q}$ dependence is straightforward to include with a form factor, typically calculated in the dipole approximation for the electron orbitals responsible for underlying magnetism.~\cite{Brown_book_2006}  However, the present measurements are for very strongly correlated electronic states between U and Pt~\cite{Joynt_RevModPhys_2002} -- possibly requiring a substantial correction to the overall magnitude of the form factor.\cite{Zaliznyak_chapter_2005} A straightforward approach that avoids such complications is to compare our polarized neutron measurements of $M$ with SQUID magnetization measurements that inherently probe the total magnetization at q=0 and which we have performed on similar crystals.  The normal state magnetization is strongly temperature dependent particularly for $H \perp c$ as can be seen in Fig. \ref{Chi_ab}(a). We have measured the temperature dependence of the flipping ratio at the (1,0,0) reflection for $H =1.0$\,T from 2 K to 230 K for this orientation and find that it scales directly with our SQUID measurements.  Normalization of the raw data in the form of $1 - R$ by this constant factor gives the absolute magnetization plotted in Fig.~\ref{Chi_ab} (a) and (b). We find that the normalization needed for $H||c$, Fig.~\ref{Chi_c}, is precisely the same as that from our measurements with $H \perp c$. 

In Fig.~\ref{Chi_ab}(a), reproduced from our earlier work,~\cite{Gan2012}  we show the complete  temperature dependence of the susceptibility from neutron scattering at 1.0~T along with our SQUID measurements including earlier susceptibility measurements from Frings {\it et. al.}\cite{Frings_JMMM_1983} They are all in excellent agreement in the normal state.  Throughout the entire temperature range the neutron scattering measurements are  consistent with the NMR Knight shift and show unambiguously that there is negligible temperature dependence of the spin susceptibility in the superconducting state.  The results for $H||c$ with the smaller crystal are shown in Fig.~\ref{Chi_c} and the agreement between NMR and polarized neutron scattering is again excellent to within statistical error. The accuracy is less for this orientation since the magnetization and Knight shifts are approximately two times smaller than for $H \perp c$.  In both cases, the total susceptibility is shown and the susceptibility determined from the Knight shifts is corrected to include  the temperature independent Van Vleck susceptibility as described in more detail by Gannon {\it et al.}~\cite{Gan2012, Gan2013}  Consequently, we conclude that the temperature independence of the spin susceptibility in the superconducting state reported by Tou {\it et al.}~\cite{Tou1996a,Tou1998b} from $^{195}$Pt NMR Knight shift reflects a bulk property of \UP.

There are two  scenarios for the interpretation of these measurements.  The first is that the spin-orbit interaction is negligible and  there is no spin-momentum locking.~\cite{Tsu2012}  In this case, the upper critical field anisotropy cannot be explained by Pauli limiting for fields  along the $c$-axis, and the  theoretical model is time reversal symmetric at odds with most experiment.  The second scenario, which is consistent with most experimental evidence and the model for odd-parity superconductivity,~\cite{Sau1994,Graf_PRB_2000} is that the temperature independence of the  susceptibility is not related to the symmetry of the superconducting state.  

We have considered other sources of temperature independent behavior of the bulk susceptibility.  For example, there are prism-plane stacking faults spread throughout our  \UP\  crystals (Supplementary Materials) apparently  responsible for anisotropic quasiparticle scattering.~\cite{Kycia_PRB_1998}  However,  quasiparticle spin-flip scattering from defects, associated with a strong spin-orbit interaction, cannot explain the temperature independence.  This mechanism~\cite{Abr1962} requires the spin-orbit scattering length to be of order the coherence length and is not possible for a superconductor in the clean limit.~\cite{Kycia_PRB_1998}   Magnetic fluctuations clearly exist throughout the bulk in the superconducting state,\cite{Aeppli_PRL_1988, Carr2017} but it is not clear what role they may play.  

More likely, given the complicated Fermi surface, there are two components to the electronic susceptibility, the dominant one being related to the hybridized local moments evolving from the 5$f$ orbitals of uranium  which overshadow a second contribution from itinerant elections responsible for superconductivity.  In this case temperature independence of the bulk susceptibility and Knight shift cannot be taken as an indication of odd-parity superconductivity.  By extension, this result might be relevant for other superconducting materials with a temperature independent Knight shift.

Research support  was provided by the U.S. Department of Energy, Office of Basic Energy Sciences, Division of Materials Sciences and Engineering, under Awards No.  DE-FG02-05ER46248 (Northwestern University), No.  DE-FG02-10ER46783 (University of Notre Dame), and No. DE-SC0012311 (Rice University), and from the Danish Council for Independent Research through DANSCATT.   We acknowledge assistance with crystal growth and characterization from J. Pollanen, C. A. Collett, A. Zimmerman, and Jia Li.  We thank Elizabeth Schemm and Aharon Kapitulnik for providing the AFM image and we are grateful to ILL for their hospitality and support and to Jim Sauls, Keenan Avers,  and Vesna Mitrovic for helpful discussions.

\end{document}